# Prediction of Alzheimer's disease-associated genes by integration of GWAS summary data and expression data

Sicheng Hao[1§], Rui Wang[1], Yu Zhang[2§], Hui Zhan[3§]

[1]College of Computer and Information Science, Northeastern University, 02215, Boston, MA, USA

[2]Department of Neurosurgery, Heilongjiang Province Land Reclamation Headquarters General Hospital, 150001, Harbin, China

[3]School of Electronic Engineering, Heilongjiang University, 150001, Harbin, China

§Corresponding author

Sicheng Hao, hao.sic@husky.neu.edu

College of Computer and Information Science, Northeastern University, 02215, Boston, MA, USA

Rui Wang, wang.rui4@husky.neu.edu

College of Computer and Information Science, Northeastern University, 02215, Boston, MA, USA

Yu Zhang, ywkwy2012@163.com

Department of Neurosurgery, Heilongjiang Province Land Reclamation Headquarters General Hospital, 150001, Harbin, China

Hui Zhan, 2000092@hlju.edu.cn

School of Electronic Engineering, Heilongjiang University, 150001, Harbin, China

## Abstract

Alzheimer's disease (AD) is the most common cause of dementia. It is the fifth-leading cause of death among elderly people. With high genetic heritability (79%), finding disease causal genes is a crucial step in find treatment for AD. Following the International Genomics of Alzheimer's Project (IGAP), many disease-associated genes have been identified; however, we don't have enough knowledge about how those disease-associated genes affect gene expression and disease-related pathways. We integrated GWAS summary data from IGAP and five different expression level data by using TWAS method and identified 15 disease causal genes under strict multiple testing (alpha < 0.05), 4 genes are newly identified; identified additional 29 potential disease causal genes under false discovery rate (alpha < 0.05), 21 of them are newly identified. Many genes we identified are also associated with some autoimmune disorder.

# Introduction

Alzheimer's disease is the most common cause of dementia which is characterized by a decline in cognitive skills that affects a person's ability to perform everyday activities. Estimated 5.4 million people in the U.S. are living with Alzheimer's disease (AD). It is the fifth-leading cause of death among those age 65 and older[1]. Although some drugs showing effectiveness to mitigate the symptoms from getting worse for a limit time, no treatment can stop the disease. Heritability for the AD was estimated up to 79% [2]. However, the current finding of AD-associated genetic variants is not enough to fully explain the AD signal pathway in sufficient detail.

During recent years, with the rapid advance of next-generation DNA sequencing, identify disease-related mutation from large data set and develop treatment become possible [3-5]. Genome-wide comparison studies (GWAS) have identified a significant amount of common genetic variants associated with complex traits and diseases [6-8]. Many previous studies have identified genes such as APOE[9, 10] on chromosome 19. However, the causal relation of those associated genes and variants remain unclear. For example, recent study and data showed that a female with the APOE gene under greater risk than a male with the APOE gene[11, 12]. This strongly indicates that we have little knowledge about how this risk factor effect people.

With GWAS summary data provided by the International Genomics of Alzheimer's Project (IGAP) [13], we are able to study AD in great detail. For a complex disease such as AD, the top single nucleotide polymorphisms (SNPs) often located in the noncoding region, hard to know which gene is modified by that mutation and many significant SNPs are in high linkage disequilibrium with non-significant SNPs, plus many associated SNPs are more likely to locate in expression regulation region of the disease causal gene[14, 15]. To identify disease-associated genes, we used the transcriptome-wide association study (TWAS) [16] method which integrates GWAS summarization level data, expression level data from human tissue. TWAS method can eliminate potential confounding and find disease causal gene by focusing only on expression trait linking related by genetic variation; it can also increase statistical power from the lower multiple-testing burden and the noise reduction of gene expression from environmental factors[16].

Previous studies have pointed out at AD is closely related to autoimmune disorders. [17, 18]. After detecting possible disease causal gene for AD, we manually curated existing research about the autoimmune diseases that potentially related to AD.

# Data and Method

Data we used for SNP-trait association is a large-scale GWAS summary data provided by IGAP with total 17,008 AD cases and 37,154 controls, include 7,055,881 SNPs, we selected 6,004,159 SNPs. Expression level data are from adipose tissue (RNA-seq), whole blood (RNA array), peripheral blood (RNA array), brain tissue (RNA-seq and RNA-seq splicing) [19-22].

### Transcriptome-wide Association Study

TWAS can be viewed as a test for correlation between predicted gene expression and traits from GWAS summary association data. The predicted effect size of gene expression on traits can be viewed as a linear model of genotypes with weights based on the correlation between SNPs and gene expression in the training data while accounting for linkage disequilibrium (LD) among SNPs.

There are eight modes of causality for the relationship between genetic variant, gene expression, and traits. Scenarios E-H in Figure-1 should be identified as significant by TWAS and its corresponding null hypothesis is gene expression completely independent of traits (A-D in Figure 1). By only focusing on the genetic component of expression, the instances of expression-trait association that is not caused by genetic variation but variation in traits can be avoided. One aspect that needs to be noticed is, same as other methods, TWAS is also confounded by linkage and pleiotropy.

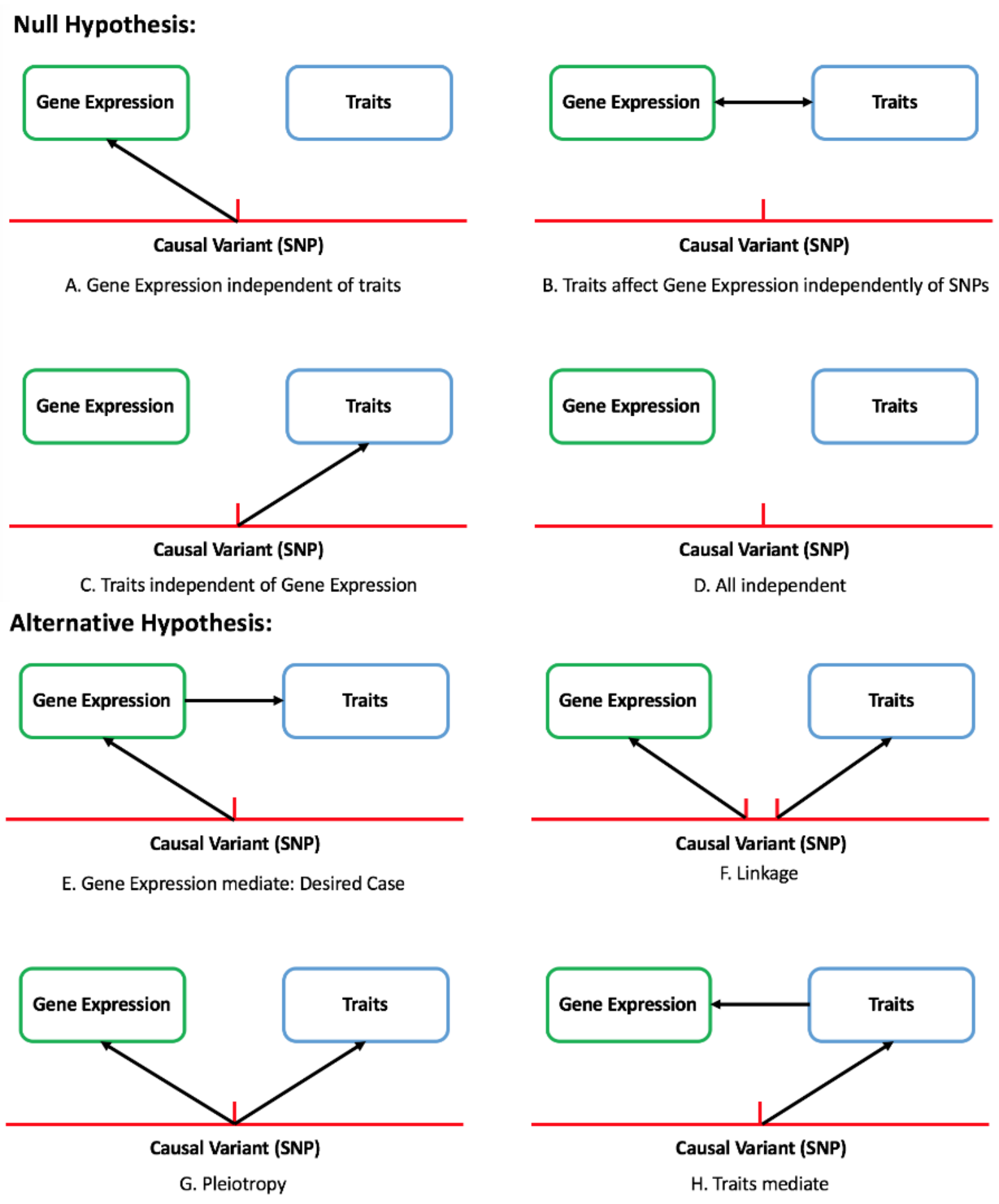

**Figure-1**

**Eight scenarios of causal assumption between gene, expression and trait in TWAS study. Null hypothesis: gene expression is completely independent of traits(A-D) Alternative hypothesis: causal relation exists between SNPs and traits(E-H)**

## Performing TWAS with GWAS Summary Statistics

We integrated gene expression measurements from 5 tissues with summary GWAS to perform multi-tissue transcriptome-wide association. In each tissue, TWAS used cross-validation to compare predictions from the best cis-eQTL to those from all SNPs at the locus. Prediction models choosing from BLUP, BSLLM [23], LASSO, and elastic net [24].

TWAS Imputes effect size ($z$-score) of the expression and trait are linear combination of elements of $z$-score of SNPs for traits with weights. The weights, $W = \Sigma_{e,s}\Sigma_{s,s}^{-1}$, are calculated using ImpG-Summary algorithm[25] and adjusted for Linkage Disequilibrium(LD) . $\Sigma_{e,s}$ is the estimated covariance matrix between all SNPs at the locus and gene expression and $\Sigma_{s,s}$ is the estimated covariance among all SNPs which is used to account for LD.

Standardized effect sizes (Z-scores) of SNPs for a trait at a given cis locus can be denoted as a vector $Z$. Also, the imputed Z-score of expression and trait, $WZ$, has variance. $W\Sigma_{s,s}W^t$. Therefore, the imputation $Z$ score of the cis genetic effect on the trait is,

$$Z_{TWAS} = WZ/(W\Sigma_{s,s}W^t)^{1/2}.$$

Bonferroni correction is usually applied when identifying significant disease-associated gene. The standard multiple testing conducted in TWAS is 0.05/15000 [16]. But traditional P-value cutoffs adjusted by Bonferroni correction are made too strict in order to avoid an abundance of false positive results. The thresholds like 0.05/15000 for significant genes are usually chosen so that the probability of any single false positive among all loci tested is smaller than 0.05, which will lead to many missed findings. Instead, False Discovery Rate error measure is a more useful approach when a study involves a large number of tests, since it can identify as many significant genes as possible while incurring a relatively low proportion of false positives.[26] For each tissue, we used the Benjamini-Hochberg procedure [27] in addition to the Bonferroni correction for all gene tested. The Benjamini-Hochberg procedure is one of false discovery rate procedures that are designed to control the expected proportion of false positives. It is less stringent than the Bonferroni correction, thus has greater power. Since this is study is more exploratory, we can pay more risk of type I error for larger statistical power. It works as follows:

Put individual p-values in ascending order and assign ranks to the p-values.

1. Calculate each individual Benjamini-Hochberg critical value with formula $\frac{k}{m}\alpha$, where $k$ is individual p-value's rank, $m$ is total number of tests and $\alpha$ is the false discovery rate.
2. Find the largest $k$ such that $P_k \leq \frac{k}{m}\alpha$ and reject the null hypothesis for all $H_i$ for $i = 1, \dots k$.

# Results

To determine which gene is significantly associated with AD, we first performed strict multiple testing Bonferroni correction (P-value < 0.05/15000). (Sample results see Figure-2) We found 15 significant genes (Table-1), 11 of them has identified by previous studies of AD.  In order to increase the search range, we performed false discovery rate under the same alpha (0.05). After the Benjamini-Hochberg procedure [27], we found 29 additional genes (Table-2). Nine of those genes has previously identified to be related to AD.

PVRL2(P-value 4.92*10^-34 in Brain (CMC) RNA-seq, also known as NECTIN2) is a well-known gene for AD. This gene encodes a single-pass type I membrane glycoprotein and interact with AOPE gene [28]. TOMM40 (P-value 1.13*10^-25 in Whole Blood (YFS) RNA Array) is also located adjacent to APOE. It has been identified by previous studies worldwide as AD related gene [29-31]. It is the central and essential component of the translocase of the outer mitochondrial membrane[32]. This confirmed that mitochondrial dysfunction plays a significant role in AD-related pathology[33, 34]

Other highly connected genes function group identified are BIN1(P-value $1.18 \times 10^{-6}$in Whole Blood (YFS) RNA Array; $3.24 \times 10^{-6}$in Peripheral Blood(NTR) RNA Array), CLU(P-value$1.45 \times 10^{-16}$, MS4A6A(P-value $5.72 \times 10^{-8}$in Peripheral Blood(NTR) RNA Array; $2.92 \times 10^{-10}$in Whole Blood (YFS) RNA Array)[35].

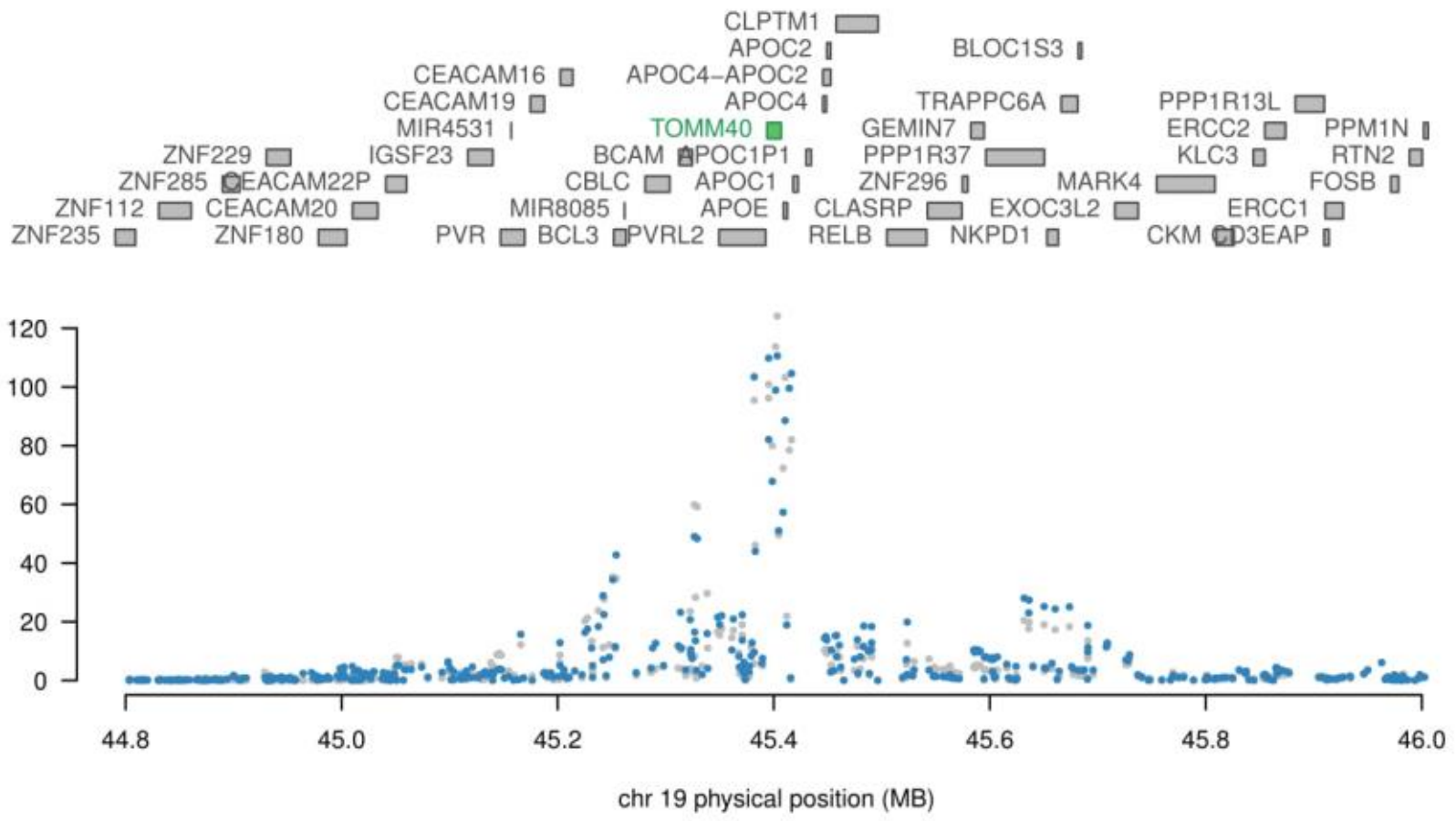

**Figure-2**

**Gene position plot in chromosome 19.**

### New Identified Genes

MLH3(P-value $7.86 \times 10^{-9}$ in Brain (CMC) RNA-seq splicing) FNBP4(P-value $1.49 \times 10^{-6}$in Whole Blood (YFS) RNA Array, CEACAM19(P-value $3.38 \times 10^{-11}$ in Adipose (METSIM) RNA-seq), and CLPTM1(P-value $5.73 \times 10^{-17}$ in Brain (CMC) RNA-seq) are newly identified AD-associated genes. MLH3 gene is known for its function in repair mismatched DNA and risk for thyroid cancer and lupus[36-38]. CEACAM19 gene located in chromosome 19, a previous study showed high expression of CEACAM19 for patients with breast cancer[39]; CLPTM1 has been shown to increase the risk of lung cancer and melanoma [40, 41]. Both CEACAM19 and CLPTM1 gene are located in chromosome 19 and near APOE gene. More detailed studies are needed to investigate the relationship between those genes and whether CLPTM1 and CEACAM19 are disease causal gene.

# Discussion

### APOE Related Genes

Although APOE is not reported to be significant in any tissue, not enough evidence to conclude that APOE is not related to AD. Since each SNP has a weight assigned regarding the expression in TWAS study, even two genes are both significantly related to a disease, it is very likely only one of them will be showing significant in TWAS. TOMM40(P-value $1.13 \times 10^{-25}$) gene located adjacent to APOE[42], and has a strong linkage disequilibrium with APOE gene[43], hence TWAS didn't detect this APOE does not imply it is not disease causal gene.  APOE and TOMM40 may interact to affect AD pathology such as mitochondrial dysfunction [44, 45]. Further study is needed to show causal relation in detail. PICALM (P-value $2084 \times 10^{-7}$ in Peripheral Blood (NTR) RNA Array) and PTK2B(P-value $9.93 \times 10^{-8}$ in Peripheral Blood (NTR) RNA Array; P-value $2.89 \times 10^{-6}$ in Whole Blood (YFS) RNA Array) are also related to APOE and TOMM40 gene according to previous studies [35, 46-48].

### Association with Autoimmune Diseases

Complex disease such as AD, often shares common pathways or causal genes with other diseases.[49]  For instance, TOMM40 is a shared disease-associated gene between AD and Type II diabetes[50]. Recent studies showing autoimmune diseases have closed relation with AD [51-53]. Among all the genes we identified through TWAS method, eight of them are related to autoimmune diseases.

As shown in Figure-3, PICALM, PVRL2, PVR, and CLU have shown to be related to systemic, an autoimmune disease characterized by vascular injury and debilitating tissue fibrosis [54-57]. CR1 and CLU gene are related to thymus function which could potentially cause an autoimmune disorder [58, 59]. MLH3 and BIN1 gene have shown to be associated with Lupus, another severe autoimmune disease [37, 60]. Although with existing result, we don't have enough evidence to prove these genes are both disease causal genes for AD and autoimmune disease, further research from areas such as metabolomics and proteomics is needed to study the disease association between AD and autoimmune diseases [61-63].

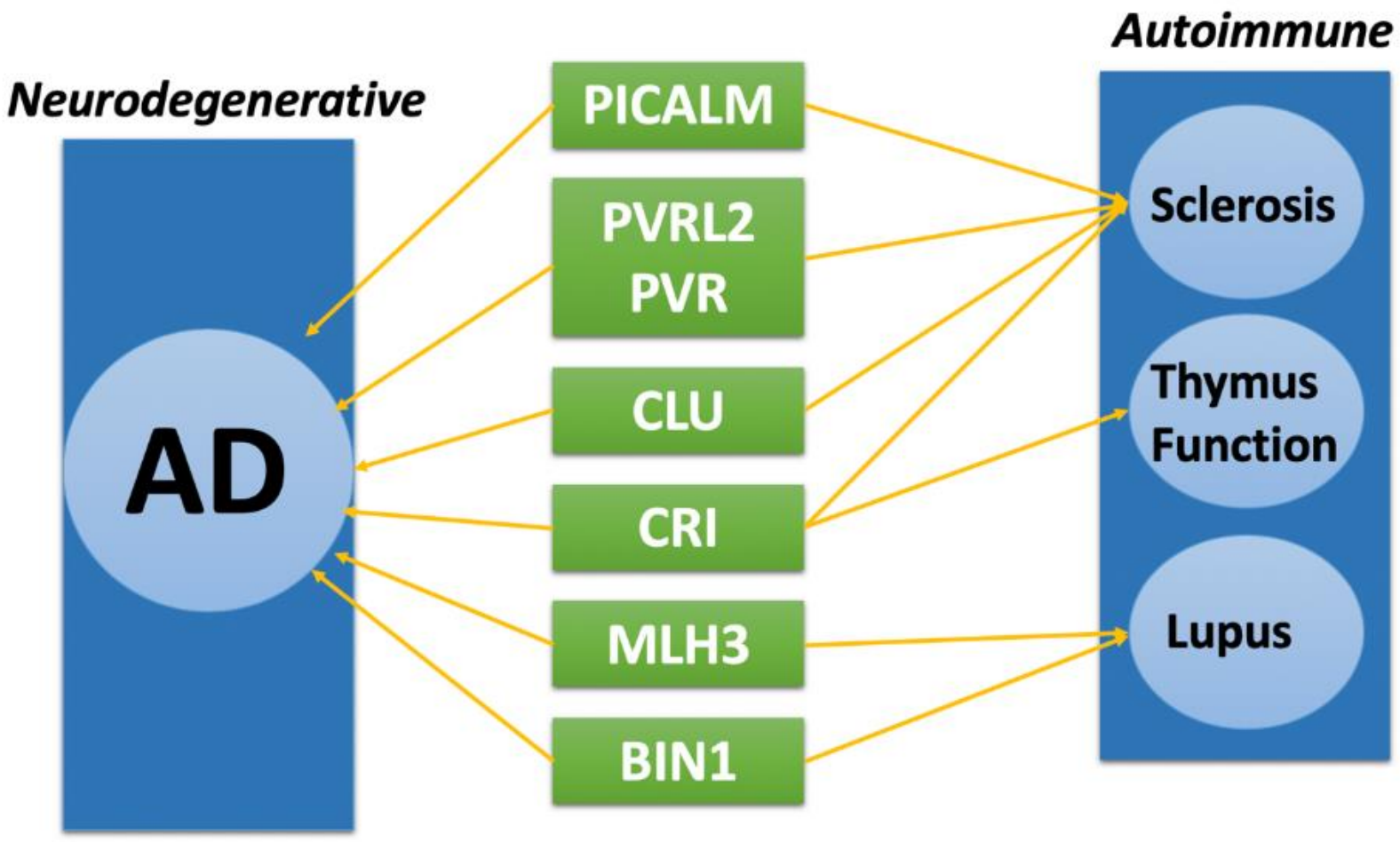

**Figure-3**

**Shared disease associated gene between Alzheimer’s disease and Autoimmune diseases**

# Appendix

**Table1: Significant genes identified by TWAS under strict multiple testing**

| Gene | Chromosome | Tissue | P-value | Z-score | Related to Autoimmune Diseases |
|---|---|---|---|---|---|
| PVRL2 | 19 | Brain (CMC) RNA-seq | 4.92E-34 | -12.1626 | Yes |
| TOMM40 | 19 | Whole Blood (YFS) RNA Arr ay | 1.13E-25 | 10.4749 | |
| CLPTM1 | 19 | Brain (CMC) RNA-seq | 5.73E-17 | -8.37061 | |
| CLU | 8 | Brain (CMC) RNA-seq splicing | 1.45E-16 | -8.26075 | |
| CR1 | 1 | Brain (CMC) RNA-seq | 4.08E-15 | 7.8523 | Yes |
| CEACAM19 | 19 | Adipose (METSIM) RNA-seq | 3.38E-11 | 6.62905 | Yes |
| MS4A6A | 11 | Whole Blood (YFS) RNA Array | 2.92E-10 | 6.30316 | |
| TRPC4AP | 20 | Brain (CMC) RNA-seq splicing | 9.43E-10 | 6.1188 | |
| MLH3 | 14 | Brain (CMC) RNA-seq splicing | 7.86E-09 | -5.77148 | Yes |
| MS4A6A | 11 | Peripheral Blood (NTR) RNA Array | 5.72E-08 | 5.4272 | |
| PTK2B | 8 | Peripheral Blood (NTR) RNA Array | 9.93E-08 | 5.32809 | |
| PVR | 19 | Brain (CMC) RNA-seq | 2.05E-07 | -5.19443 | Yes |
| PICALM | 11 | Peripheral Blood (NTR) RNA Array | 2.84E-07 | 5.1337 | Yes |
| MS4A4A | 11 | Adipose (METSIM) RNA-seq | 6.11E-07 | 4.99 | |
| BIN1 | 2 | Whole Blood (YFS) RNA Array | 1.18E-06 | 4.859114 | |
| FNBP4 | 11 | Whole Blood (YFS) RNA Array | 1.49E-06 | -4.81307 | |
| PTK2B | 8 | Whole Blood (YFS) RNA Array | 2.89E-06 | 4.6784 | Yes |
| BIN1 | 2 | Peripheral Blood (NTR) RNA Array | 3.24E-06 | 4.65503 | Yes |

**Tabel2:Additional gene under Benjamini-Hochberg procedure**

| Gene | Chromosome | Tissue | P-value | Z-score | Previously Identified |
|---|---|---|---|---|---|
| PHACTR1 | 6 | Whole Blood (YFS) RNA Array | 3.41E-06 | -4.64434 | |
| PTPMT1 | 11 | Whole Blood (YFS) RNA Array | 4.45E-06 | 4.58895 | |
| MTCH2 | 11 | Peripheral Blood (NTR) RNA Array | 5.76E-06 | 4.535 | |
| C1QTNF4 | 11 | Adipose (METSIM) RNA-seq | 8.82E-06 | 4.44 | |
| FAM180B | 11 | Brain (CMC) RNA-seq | 1.09E-05 | -4.39814 | Yes |
| DMWD | 19 | Whole Blood (YFS) RNA Array | 1.22E-05 | 4.3733 | |
| ELL | 19 | Whole Blood (YFS) RNA Array | 1.89E-05 | 4.277 | Yes |
| ZNF740 | 12 | Brain (CMC) RNA-seq splicing | 2.08E-05 | 4.25599 | |
| NYAP1 | 7 | Adipose (METSIM) RNA-seq | 2.47E-05 | -4.21777 | |
| SDAD1 | 4 | Whole Blood (YFS) RNA Array | 3.04E-05 | -4.17062 | |
| MTSS1L | 16 | Brain (CMC) RNA-seq splicing | 3.35E-05 | 4.14833 | |
| PHKB | 16 | Brain (CMC) RNA-seq | 3.70E-05 | -4.1257 | Yes |
| SLC39A13 | 11 | Brain (CMC) RNA-seq splicing | 4.01E-05 | -4.10667 | Yes |
| CD33 | 19 | Whole Blood (YFS) RNA Array | 4.04E-05 | 4.1051 | Yes |
| AP2A2 | 11 | Brain (CMC) RNA-seq | 4.28E-05 | -4.09193 | Yes |
| ZYX | 7 | Adipose (METSIM) RNA-seq | 4.56E-05 | -4.07718 | |
| ZNF232 | 17 | Brain (CMC) RNA-seq splicing | 4.73E-05 | -4.0688 | |
| ZNF232 | 17 | Brain (CMC) RNA-seq splicing | 4.76E-05 | 4.0671 | |
| DLST | 14 | Peripheral Blood (NTR) RNA Array | 5.26E-05 | 4.0436 | Yes |
| TBC1D7 | 6 | Adipose (METSIM) RNA-seq | 5.34E-05 | 4.0403 | |
| ELL | 19 | Adipose (METSIM) RNA-seq | 5.48E-05 | 4.03401 | |
| SLC39A13 | 11 | Brain (CMC) RNA-seq splicing | 5.79E-05 | -4.02128 | Yes |
| TMCO6 | 5 | Whole Blood (YFS) RNA Array | 6.50E-05 | 3.9938 | |
| CEL | 9 | Whole Blood (YFS) RNA Array | 6.99E-05 | 3.97671 | Yes |
| MYBPC3 | 11 | Adipose (METSIM) RNA-seq | 7.05E-05 | 3.97 | Yes |
| TBC1D7 | 6 | Brain (CMC) RNA-seq splicing | 7.48E-05 | -3.96063 | |
| LRRC25 | 19 | Peripheral Blood (NTR) RNA Array | 7.74E-05 | -3.9523 | |
| TBC1D7 | 6 | Brain (CMC) RNA-seq splicing | 8.37E-05 | 3.93351 | |
| KIR3DX1 | 19 | Peripheral Blood (NTR) RNA Array | 8.87E-05 | 3.9195 | |
| SIX5 | 19 | Peripheral Blood (NTR) RNA Array | 9.32E-05 | 3.9076 | |
| HBEGF | 5 | Whole Blood (YFS) RNA Array | 9.92E-05 | -3.8926 | Yes |
| NUP88 | 17 | Peripheral Blood (NTR) RNA Array | 1.60E-04 | -3.7748 | |
| FAM105B | 5 | Whole Blood (YFS) RNA Array | 1.61E-04 | 3.773 | |
| ARL6IP4 | 12 | Peripheral Blood (NTR) RNA Array | 2.10E-04 | 3.707 | |